\documentclass[prd,preprint,a4paper,superscriptaddress,nofootinbib,11pt]{revtex4}

\usepackage{amsmath,amsfonts,amssymb}
\usepackage[T1]{fontenc} 
\usepackage{euscript}

\usepackage[usenames, dvipsnames]{color}

\DeclareMathAlphabet{\mathpzc}{T1}{pzc}{m}{it}

\newcommand{\kl}{l} 

\begin{document}

\begin{center}
{\bf  \Large Realizations of $\kappa$-Minkowski space, Drinfeld twists and related symmetry algebras}
 
 \bigskip
\bigskip

Tajron Juri\'c, \footnote{e-mail: tjuric@irb.hr} Stjepan Meljanac, \footnote{e-mail: meljanac@irb.hr}  Danijel Pikuti\'c, \footnote{e-mail: dpikutic@irb.hr} \\  
Ru\dj er Bo\v skovi\'c Institute, Theoretical Physics Division, Bijeni\v cka  c.54, HR-10002 Zagreb,
Croatia \\[3mm]

\end{center}
\setcounter{page}{1}


\vspace{1.0cm}
{
Realizations of $\kappa$-Minkowski space linear in momenta are studied for time-, space- and light-like deformations. We construct and classify all such linear realizations and express them in terms of $\mathfrak{gl}(n)$ generators. There are three one-parameter families of linear realizations for time-like and space-like deformations, while for light-like deformations, there are only four linear realizations. The relation between deformed Heisenberg algebra, star product, coproduct of momenta and twist operator is presented. It is proved that for each linear realization there exists Drinfeld twist satisfying normalization and cocycle conditions. $\kappa$-deformed $\mathfrak{igl}(n)$-Hopf algebras are presented for all cases. The $\kappa$-Poincar\'e-Weyl and $\kappa$-Poincar\'e-Hopf algebras are discussed. Left-right dual $\kappa$-Minkowski algebra is constructed from the transposed twists. The corresponding realizations are nonlinear. All Drinfeld twists related to $\kappa$-Minkowski space are obtained from our construction. Finally, some physical applications are discussed. 
}



\newpage
\tableofcontents

\section{Introduction}

One of the biggest problems in fundamental theoretical physics is a great difficulty to reconcile quantum mechanics and general theory of relativity in order to formulate consistent theory of quantum gravity.  It is argued that at very high energies the gravitational effects can no longer be neglected and that the spacetime is no longer a smooth manifold, but rather a fuzzy, or better to say a noncommutative space \cite{doplicher}. Physical theories on such noncommutative manifolds require a new framework. This new framework is provided by noncommutative geometry \cite{connes}. In this framework, the search for generalized (quantum) symmetries that leave the physical action invariant leads to deformation of Poincar\'e symmetry, with $\kappa$-Poincar\'e symmetry being among the most extensively studied \cite{Lukierski, Majid, Kowalski, 12}.

$\kappa$-deformed Poincar\'e symmetry is algebraically described by the $\kappa$-Poincar\'e-Hopf algebra and is an example of deformed relativistic symmetry that can possibly describe the physical reality at the Planck scale. $\kappa$ is the deformation parameter usually interpreted as the Planck mass or some quantum gravity scale. It was shown that  quantum field theory with $\kappa$-Poincar\'e  symmetry emerges in a certain limit of quantum gravity coupled to matter fields after integrating out the gravitational/topological degrees of freedom \cite{topology}. This amounts to an effective theory in the form of a noncommutative field theory on the $\kappa$-deformed Minkowski space. 

It is known \cite{8} that the deformations of the symmetry group  can be realized through the application of the Drinfeld twist on that symmetry group \cite{Drinfeld, Drinfeld1, alltheway, NCSpacetimes}.  The main virtue of the twist formulation is that the deformed (twisted) symmetry algebra is the same as the original undeformed one and that there is only a change in the coalgebra structure which then leads to the same free field structure as the corresponding commutative field theory.

In \cite{BorLukPach} it was shown that the coproduct of $D=2$ and $D=4$ quantum $\kappa$-Poincar\'e algebra in the classical basis can not be obtained by the cochain twist  depending only on the Poincar\'e generators (even if the coassociativity condition is relaxed). However the deformation used in \cite{BorLukPach} is the so called time-like type of deformation, and it is known \cite{51c} that for light-like deformation such a twist indeed exists \cite{KulishMudrov, BorPach, universal, kratki}.

In this work, we go other way round. Starting from $\kappa$-Minkowski space, we obtain its linear realizations, then coproducts of momenta from realizations and finally, we present a method for obtaining corresponding twists from those coproducts. We show that, for linear realizations, those twists are Drinfeld twists, satisfying normalization and cocycle conditions. The method for obtaining  Drinfeld twists corresponding to each linear realization is elaborated and it is shown how these twists generate new Hopf algebras.
The resulting symmetry algebras are $\kappa$-deformed $\mathfrak{igl}(n)$ Hopf algebras. In special cases we obtain $\kappa$-Poincar\'e-Weyl-Hopf algebra and $\kappa$-Poincar\'e-Hopf algebra, but the former is obtained only for the case of light-like deformation.

The paper is organized as follows. In the second section, $\kappa$-Minkowski spacetime with deformation vector $a_\mu$ in various directions (timelike, spacelike and lightlike) is introduced. In section III, notion of linear realizations is introduced and all linear realizations in $n$ dimensions for $n>2$ are found. Those realizations are then expressed in terms of generators of $\mathfrak{gl}(n)$ algebra. In section IV, deformed Heisenberg algebra is presented, along with star product and coproducts of momenta. At the end of this section, the twist operator is introduced and the relation between star product, twist operator and coproduct of momenta is given. In section V it is shown that the twist operator from the previous section is a Drinfeld twist, satisfying normalization and cocycle conditions. It is shown that initial linear realizations follow from these twists, which confirms the consistency of our approach. At the end of the section V, $\mathcal R$-matrix is presented. In the section VI, $\kappa$-deformed $\mathfrak{igl}(n)$ Hopf algebra is presented, in general and for four special cases. In section VII, left-right dual $\kappa$-Minkowski algebra is constructed from the transposed twists. Alternatively, $\kappa$-Minkowski algebra is obtained from transposed twists with $a_\mu \rightarrow - a_\mu$. The corresponding realizations are non-linear. In section VIII, nonlinear realizations of $\kappa$-Minkowski space and related Drinfeld twists, known in the literature so far, are presented. Finally, in section IX, outlook and discussion are given.

\section{$\kappa$-Minkowski space}

$\kappa$-Minkowski space is usually defined by \cite{Lukierski, Majid, Kowalski, Zarkzewski}:
\begin{equation}
[\hat x_0, \hat x_i] = \frac i\kappa \hat x_i, \qquad [\hat x_i, \hat x_j] = 0.
\label{standardkm}
\end{equation}
Equations \eqref{standardkm} can be rewritten in a covariant way \cite{LL96}:
\begin{equation}
[\hat x_\mu, \hat x_\nu] = i(a_\mu \hat x_\nu - a_\nu \hat x_\mu)
\label{commx}
\end{equation}
where $a_\mu \in \mathbb M^n$ ($\mathbb M^n$ being undeformed $n$-dimensional Minkowski space) is a fixed deformation vector, which for the choice $a_0=\kappa^{-1}$ and $a_i=0$ corresponds to \eqref{standardkm}. Noncommutative coordinates $\hat x_\mu$ of $\kappa$-Minkowski space form a Lie algebra.

Note that Lie algebra \eqref{commx} is independent of metric. However, we point out that our physical requirement is that in the limit $a_\mu\rightarrow0$, we get ordinary Minkowski spacetime. Hence, it is natural to assume and treat $a_\mu$ as a vector in undeformed Minkowski space. There are two possibilities. One is to fix real parameters $a_\mu$ and the other is when $a_\mu$ are not fixed (transforming together with noncommutative coordinates $\hat x_\mu$) \cite{andelo1007}. In this paper we chose the first possibility.

Throughout the article, indices are raised and lowered by the Minkowski metric $\eta$, i.e. $a^\mu=\eta^{\mu\nu}a_\nu$ and $a_\mu=\eta_{\mu\nu}a^\nu$, where the convention with positive spatial eigenvalues of the metric is used, e.g. in $(3+1)$ dimensions $\eta=\mathrm{diag}(-1,1,1,1)$. Also, indices are contracted the same way, i.e. $a\cdot b=a_\mu b^\mu=\eta^{\mu\nu}a_\mu b_\nu$ and 
$a^2 = a\cdot a = a_\mu a^\mu=\eta^{\mu\nu}a_\mu a_\nu$ for any vectors $a_\mu$ and $b_\mu$.

The deformation vector can be timelike ($a^2<0$), lightlike ($a^2=0$) and spacelike ($a^2>0$), so it can be written like:
\begin{equation}
a_\mu = \frac{1}{\kappa} u_\mu
\end{equation}
where $\kappa^{-1}$ is expansion parameter and $u^2 \in \{-1,0,1\}$, which corresponds to the previously mentioned three cases. Light-like deformation $a^2=0$ was first treated in context of null-plane quantum Poincar\'e algebra \cite{balesteros}. 
Depending on the sign of $a^2$, $\kappa$-Minkowski Lie algebra is invariant under the following little groups:
\begin{itemize}
\item If $a_\mu$ is timelike ($a^2<0$), the little group is $SO(n-1)$
\item If $a_\mu$ is lightlike ($a^2=0$), the little group is $E(n-2)$
\item If $a_\mu$ is spacelike ($a^2>0$), the little group is $SO(n-2,1)$
\end{itemize}

It is useful to introduce enveloping algebra $\hat{\mathcal A}$, generated by the elements $\hat x_\mu$ of $\kappa$-Minkowski algebra.

\section{Linear realizations}
Commutative coordinates $x_\mu$ and momenta $p_\mu$ generate an undeformed Heisenberg algebra $\mathcal H$ given by:
\begin{equation}\begin{split}\label{undefHA}
&[x_\mu,x_\nu]=0,  \\
&[p_\mu, x_\nu]=-i\eta_{\mu\nu}, \\
&[p_\mu,p_\nu]=0
\end{split}\end{equation}
Analogously to $\hat{\mathcal A}$ in the previous section, commutative coordinates $x_\mu$ generate enveloping algebra $\mathcal A$, which is subalgebra of undeformed Heisenberg algebra, i.e. $\mathcal A \subset \mathcal H$. Momenta $p_\mu$ generate algebra $\mathcal T$, which is also subalgebra of undeformed Heisenberg algebra, i.e. $\mathcal T \subset \mathcal H$. Undeformed Heisenberg algebra is, symbolically, $\mathcal H = \mathcal A \mathcal T$.

In general, realization of NC space is given by:
\begin{equation} \label{generic-realization}
\hat x_\mu = x_\alpha {\varphi^\alpha}_\mu(p)
\end{equation}
where ${\varphi^\alpha}_\mu(p)$ is a function of $p_\mu$ which should reduce to $\delta^\alpha_\mu$ in the limit when deformation goes to zero 
\cite{Meljanac-3, Meljanac-4, Dimitrijevic}.

It is important to note that different realizations $\hat x_\mu=x_\alpha {\varphi^\alpha}_\mu(p)=x'_\alpha {\varphi'^\alpha}_\mu(p')$ are related by similarity transformations, where $(x_\mu, p_\nu)$ and $(x'_\mu, p'_\nu)$ satisfy undeformed Heisenberg algebra \cite{sasa0804}\cite{algebroid}.
In this section, additional label for $x_\mu$ and $p_\mu$ is omitted for the sake of simplicity.

We are looking for linear realizations of $\kappa$-Minkowski space, that is the realizations where the function ${\varphi^\alpha}_\mu(p)$ is linear in $p_\mu$. They can be written in the form
\begin{equation}
\hat x_\mu = x_\mu + \kl_\mu
\label{xplusm}
\end{equation}
where $\kl_\mu$ is linear in momentum $p_\mu$. It is given by:
\begin{equation}
\kl_\mu = {K_{\beta\mu}}^\alpha x_\alpha p^\beta
\label{mKxp}
\end{equation}
where ${K_{\beta\mu}}^\alpha \in \mathbb R$. Inserting it in (\ref{commx}) gives that ${K_{\mu\nu}}^\alpha$ has to satisfy:
\begin{gather}
{K_{\mu\nu}}^\alpha-{K_{\nu\mu}}^\alpha=a_\mu \delta_\nu^\alpha - a_\nu \delta_\mu^\alpha
\label{KisC} \\
{K_{\gamma\mu}}^\alpha {K_{\beta\nu}}^\gamma-
{K_{\gamma\nu}}^\alpha {K_{\beta\mu}}^\gamma
=
a_\mu{K_{\beta\nu}}^\alpha-
a_\nu{K_{\beta\mu}}^\alpha
\label{KKCK}
\end{gather}
It also follows that $\kl_\mu$ satisfies the same commutation relations as $\hat x_\mu$:
\begin{equation}
[\kl_\mu, \kl_\nu]=i(a_\mu \kl_\nu - a_\nu \kl_\mu)
\label{commm}
\end{equation}


\subsection{Classification of linear realizations}

Since we assume that equations \eqref{KisC} and \eqref{KKCK} transform under Lorentz algebra, the most general covariant ansatz for ${K_{\mu\nu}}^\alpha$ in terms of deformation vector $a_\mu$ for arbitrary number of dimensions\footnote{In 2 dimensions there are additional terms constructed with two dimensional Levi-Civita tensor $\epsilon_{\mu\nu}$. For example, there is a solution $K_{\mu\nu\lambda}=\frac{a_\mu a_\nu}{a^2}(c_1 a_\alpha + c_2 \epsilon_{\alpha\beta}a^\beta)$, where $c_1, c_2 \in \mathbb R$ are parameters and $a^2\ne0$.} $n>2$ is:
\begin{equation}
{K_{\mu\nu}}^\alpha = A_0 a_\mu a_\nu a^\alpha + A_1 \eta_{\mu\nu} a^\alpha + A_2 \delta_\mu^\alpha a_\nu + A_3 a_\mu \delta_\nu^\alpha
\label{kaminia}
\end{equation}
leading to the following $\kl_\mu$:
\begin{equation}
\kl_\mu = A_0 a_\mu(a\cdot x)(a\cdot p)+
A_1(a\cdot x)p_\mu+A_2a_\mu(x\cdot p)+A_3x_\mu(a\cdot p)
\end{equation}

From equation (\ref{KisC}) it follows that 
\begin{equation}
A_3 = A_2+1. 
\label{aa+1}
\end{equation}
Using (\ref{KKCK}) in combination with \eqref{aa+1} yields the following equations:
\begin{align}
&A_1(A_0a^2+A_1+1)=0 \\
&A_3(A_0a^2+A_3+1)=0 \\
&A_1A_3a^2=0
\end{align}

Those equations have four solutions:
\begin{enumerate}
\item $A_1=0, ~~~~ A_2=-1, ~~~~ A_3=0, ~~~~ a^2A_0=c$
\item $A_1=0, ~~~~ A_2=-c, ~~~~ A_3=1-c, ~~~~ a^2A_0=c$
\item $A_1=-1-c, ~~~~ A_2=-1, ~~~~ A_3=0, ~~~~ a^2A_0=c$
\item $A_1=-1, ~~~~ A_2=0, ~~~~ A_3=1, ~~~~ a^2=A_0=0$
\end{enumerate}
where $c\in\mathbb R$ is a free parameter. We will denote these  four types of realizations by $\mathcal C_1$, $\mathcal C_2$, $\mathcal C_3$ and $\mathcal C_4$ respectively\footnote{Where $\mathcal C$ stands for \emph{covariant}.}.  Explicitly for the tensor $K_{\mu\nu\alpha}$ we have
\begin{equation}\begin{split}\label{cetri}
\mathcal C_1 :&\quad K_{\mu\nu\alpha}= \begin{cases}
\dfrac c{a^2}a_\mu a_\nu a_\alpha - \eta_{\mu\alpha}a_\nu, & \text{if }a^2\ne 0
\\ -  \eta_{\mu\alpha}a_\nu, & \text{if }a^2=0, \end{cases} \\
\mathcal C_2 :&\quad K_{\mu\nu\alpha}=\begin{cases}
\dfrac c{a^2} a_\mu a_\nu a_\alpha-c\eta_{\mu\alpha}a_\nu+(1-c)\eta_{\nu\alpha}a_\mu,& \text{if } a^2\ne 0\\
\eta_{\nu\alpha}a_\mu,& \text{if }a^2=0, \end{cases} \\
\mathcal C_3:&\quad K_{\mu\nu\alpha}=\begin{cases}
\dfrac c{a^2} a_\mu a_\nu a_\alpha - (1+c)\eta_{\mu\nu}a_\alpha-\eta_{\mu\alpha}a_\nu, & \text{if }a^2\ne 0\\
-\eta_{\mu\nu}a_\alpha-\eta_{\mu\alpha}a_\nu, & \text{if }a^2= 0, \end{cases} \\
\mathcal{C}_{4}:&\quad K_{\mu\nu\alpha}=-\eta_{\mu\nu}a_{\alpha}+\eta_{\nu\alpha}a_{\mu}, \quad \text{only for }a^2=0.
\end{split}\end{equation}
Inserting \eqref{cetri} into \eqref{xplusm} and \eqref{mKxp} gives:
\begin{equation}\begin{split}\label{realizations}
\mathcal C_1:&\quad \hat x_\mu = \begin{cases}
x_\mu + a_\mu\left[\dfrac c{a^2} (a\cdot x)(a\cdot p) - (x\cdot p) \right],& a^2\ne 0 \\
x_\mu - a_\mu(x\cdot p),& a^2= 0,
\end{cases} \\
\mathcal C_2:&\quad \hat x_\mu = \begin{cases}
x_\mu\left[1 + (1-c)(a\cdot p)\right] + a_\mu\left[\dfrac c{a^2} (a\cdot x)(a\cdot p) - c(x\cdot p)\right],& a^2\ne 0 \\
x_\mu\left[1 + (a\cdot p)\right],& a^2= 0,
\end{cases} \\
\mathcal C_3:&\quad \hat x_\mu = \begin{cases}
x_\mu + a_\mu\left[\dfrac c{a^2}(a\cdot x)(a\cdot p) - (x\cdot p)\right] - (1+c)(a\cdot x)p_\mu,& a^2\ne 0 \\
x_\mu - a_\mu(x\cdot p) - (a\cdot x)p_\mu,& a^2= 0,
\end{cases} \\
\mathcal C_4:&\quad \hat x_\mu = x_\mu\left[1+ (a\cdot p) \right] - (a\cdot x)p_\mu, \quad \text{only for }a^2=0.
\end{split}\end{equation} 
Linear realizations for $\kappa$-deformed Euclidean space were studied in \cite{Dimitrijevic}. However, in $\kappa$-Minkowski spacetime, we have found four new linear realizations corresponding to light-like deformations ($a^2=0$). Only one of them, $\mathcal C_4$, corresponds to $\kappa$-Poincar\'e Hopf algebra \cite{universal}. 
$\mathcal C_1$ and $\mathcal C_2$ are equivalent for $c=1$ while $\mathcal C_1$ and $\mathcal C_3$ are equivalent for $c=-1$. It is important to note that the first three solutions $\mathcal C_1$, $\mathcal C_2$ and $\mathcal C_3$  are valid for all $a^2\in\mathbb R$
, and the fourth solution $\mathcal C_4$ is only valid in the case of light-like deformation.

The inverse matrices $\varphi^{-1}_{\mu\nu}$, such that ${\varphi_\mu}^\alpha \varphi^{-1}_{\alpha\nu}=\eta_{\mu\nu}$, for $\mathcal C_1$, $\mathcal C_2$, $\mathcal C_3$ and $\mathcal C_4$ are:
\begin{equation}\begin{split}
\mathcal C_1:\quad \varphi^{-1}_{\mu\nu} =&
\begin{cases}
\eta_{\mu\nu} + \dfrac1{1-(a\cdot p) +c}\left(p_\mu - c \dfrac{a_\mu}{a^2} \right)a_\nu,& a^2\ne0\\
\eta_{\mu\nu} + \dfrac{p_\mu a_\nu}{1-(a\cdot p)},& a^2=0,
\end{cases}
\\
\mathcal C_2:\quad \varphi^{-1}_{\mu\nu} =&
\begin{cases}
 \dfrac{\eta_{\mu\nu}}{1+(1-c)(a\cdot p)}
-\dfrac c{\left[1+(1-c)(a\cdot p) \right]^2}\left(\dfrac{a\cdot p}{a^2}a_\mu + p_\mu \right)a_\nu,&a^2\ne0 \\
 \dfrac{\eta_{\mu\nu}}{1+(a\cdot p)}, &a^2=0,
\end{cases}
\\
\mathcal C_3:\quad \varphi^{-1}_{\mu\nu} =&
\begin{cases}
\begin{split} &\eta_{\mu\nu} 
-\frac1{1-a\cdot p(1-c(a\cdot p))} \left[c(a\cdot p) \left(\frac{a_\mu}{a^2}+p_\mu \right) -p_\mu \right]a_\nu \\
&+\frac{(1+c)\left[(1-a\cdot p)a_\mu + a^2p_\mu \right]p_\nu}
{1-2a\cdot p + (1+c)((a\cdot p)^2 - a^2 p^2)},
\end{split}&a^2\ne0 \\
\begin{split} &\eta_{\mu\nu} 
+\frac{p_\mu a_\nu}{1-a\cdot p} +\frac{\left[(1-a\cdot p)a_\mu + a^2p_\mu \right]p_\nu}
{1-2a\cdot p + ((a\cdot p)^2 - a^2 p^2)},
\end{split}&a^2\ne0,
\end{cases}
\\
\mathcal C_4:\quad \varphi^{-1}_{\mu\nu} =&~ \frac{\eta_{\mu\nu}+a_\mu p_\nu}{1+(a\cdot p)}.
\label{inversephi}
\end{split}\end{equation}

Special cases of $\mathcal C_1$, $\mathcal C_2$ and $\mathcal C_3$, when $c=0$, we denote as $\mathcal S_1$, $\mathcal S_2$ and $\mathcal S_3$ respectively:
\begin{equation}\label{specialxp}\begin{split}
\mathcal S_1:&\quad \hat x_\mu = x_\mu - a_\mu(x\cdot p), \\
\mathcal S_2:&\quad \hat x_\mu = x_\mu\left[1 + (a\cdot p)\right], \\
\mathcal S_3:&\quad \hat x_\mu = x_\mu - a_\mu(x\cdot p) - (a\cdot x)p_\mu,
\end{split}\end{equation}
where $a^2\in \mathbb R$.



\subsection{Symmetry algebra $\mathfrak{igl}(n)$}
For fixed solution $K_{\mu\nu\lambda}$ we define undeformed $\mathfrak{igl}(n)$ algebra generated by $p_\mu$ and $L_{\mu\nu}$:
\begin{equation}\label{igln}\begin{split}
[L_{\mu\nu},L_{\lambda\rho}]&=\eta_{\nu\lambda}L_{\mu\rho}-\eta_{\mu\rho}L_{\lambda\nu} \\
[L_{\mu\nu},p_\lambda]&=-p_\nu \eta_{\mu\lambda}\\
[p_\mu,p_\nu]&=0 
\end{split}\end{equation}
In addition to commutation relations \eqref{igln}, 
\begin{equation}
[L_{\mu\nu}, x_\lambda]=x_\mu \eta_{\nu\lambda}
\end{equation} 
also holds.

Linear realizations can be written in terms of $L_{\mu\nu}$:
\begin{equation}
\hat x_\mu = x_\mu -iK_{\beta\mu\alpha}L^{\alpha\beta}
\end{equation}
Particularly, for \eqref{realizations}:
\begin{equation}\begin{split}\label{realizationsL}
\mathcal C_1:&\quad \hat x_\mu = \begin{cases}
x_\mu - ia_\mu \left(\dfrac c{a^2} a_\alpha a_\beta L^{\alpha\beta}  -{L_\alpha}^\alpha \right),&a^2\ne0 \\
x_\mu + ia_\mu {L_\alpha}^\alpha,&a^2=0,
\end{cases}\\
\mathcal C_2:&\quad \hat x_\mu = \begin{cases} 
x_\mu - i\dfrac c{a^2} a_\mu a_\alpha a_\beta L^{\alpha\beta} + ic {L_\alpha}^\alpha a_\mu 
- i(1-c)a^\alpha L_{\mu\alpha},&a^2\ne0\\
x_\mu - ia^\alpha L_{\mu\alpha},&a^2=0,
\end{cases}\\
\mathcal C_3:&\quad \hat x_\mu = \begin{cases} 
x_\mu - i\dfrac c{a^2} a_\mu a_\alpha a_\beta L^{\alpha\beta}
+ i(1+c)a^\alpha L_{\alpha\mu} + i{L_\alpha}^\alpha a_\mu,&a^2\ne0\\
x_\mu + ia^\alpha L_{\alpha\mu} + i{L_\alpha}^\alpha a_\mu,&a^2=0,
\end{cases}\\
\mathcal C_4:&\quad \hat x_\mu = x_\mu + ia^\alpha(L_{\alpha\mu}-L_{\mu\alpha})=x_\mu + ia^\alpha M_{\alpha\mu}, \quad \text{only for } a^2=0
\end{split}\end{equation}
where $M_{\mu\nu}=L_{\mu\nu}-L_{\nu\mu}$ generate Lorentz algebra. Note that $\mathcal C_4$ is the only solution that can be written in terms of Lorentz generators. 

Commutation relations between generators of $\mathfrak{igl}(n)$ algebra with $\hat x_\mu$ are
\begin{align}
[p_\mu,\hat x_\nu]&=-i\varphi_{\mu\nu}=-i(\eta_{\mu\nu}+K_{\alpha\nu\mu}p^\alpha) \label{phatx}\\
[L_{\mu\nu},\hat x_\lambda]&=\hat x_\mu \eta_{\nu\lambda}+
i(K_{\beta\lambda\alpha}\eta_{\nu\lambda}-K_{\beta\lambda\nu}\eta_{\alpha\mu}-K_{\mu\lambda\alpha}\eta_{\beta\nu})L^{\alpha\beta}
\label{Lhatx}
\end{align}
Algebra generated by $L_{\mu\nu}$, $p_\mu$ and $\hat x_\mu$, satisfies all the Jacobi relations. Only for solution $\mathcal C_4$ this is also true for algebra generated by $M_{\mu\nu}$, $p_\mu$ and $\hat x_\mu$.

At the end of this section let us introduce anti-involution operator $\dagger$ by $\lambda^\dagger=\bar\lambda$, for $\lambda\in\mathbb C$ and bar denoting the ordinary complex conjugation $(\hat x_\mu)^\dagger = \hat x_\mu$, $(x_\mu)^\dagger = x_\mu$, $(p_\mu)^\dagger = p_\mu$ and $(M_{\mu\nu})^\dagger = -M_{\mu\nu}$. Since $(a_\mu)^\dagger = a_\mu$ the relations (2), \eqref{undefHA}, \eqref{phatx} and \eqref{defHA} remain unchanged (i.e. they are invariant) under the action of $\dagger$. Note that realizations $\mathcal C_1$, $\mathcal C_2$ and $\mathcal C_3$ are generally not hermitian. In order to get the hermitian realizations, one has to make following substitutions:
 $\hat x_\mu \rightarrow \frac12(\hat x_\mu + \hat x_\mu^\dagger)$,
 $\kl_\mu \rightarrow \frac12(\kl_\mu + \kl_\mu^\dagger)$,
 $L_{\mu\nu} \rightarrow \frac12(L_{\mu\nu}-(L_{\mu\nu})^\dagger)$
 throughout the whole paper \cite{IJMPA1307}.

\section{Deformed Heisenberg algebra, star product and twist operator}

Non-commutative $\kappa$-Minkowski coordinates $\hat x_\mu$ and momenta $p_\mu$ generate a deformed Heisenberg algebra $\hat{\mathcal H}$ given by \cite{kovacevic-meljanac}:
\begin{equation}\begin{split} \label{defHA}
&[\hat x_\mu,\hat x_\nu]=i(a_\mu \hat x_\nu - a_\nu \hat x_\mu),  \\
&[p_\mu, \hat x_\nu]=-i\varphi_{\mu\nu}(p)=-i(\eta_{\mu\nu}+K_{\alpha\nu\mu}p^\alpha), \\
&[p_\mu,p_\nu]=0
\end{split}\end{equation}
From previous section, it follows that $\hat{\mathcal H}$ is isomorphic to $\mathcal H$. Algebra $\hat{\mathcal A}$ is a subalgebra of $\hat{\mathcal H}$, i.e. $\hat{\mathcal A} \subset \hat{\mathcal H}$. Deformed Heisenberg algebra is, symbolically, $\hat{\mathcal H} = \hat{\mathcal A} \mathcal T$.

\subsection{Actions $\blacktriangleright$ and $\triangleright$}

Action $\blacktriangleright$ is a map  $\blacktriangleright:  \hat{\mathcal H} \otimes \hat{\mathcal A} \rightarrow \hat{\mathcal A}$ satisfying the following properties:
\begin{gather}
\hat f \blacktriangleright \hat g = \hat f \hat g, \quad \forall \hat f, \hat g \in \hat{\mathcal A}\\
p_\mu \blacktriangleright \hat f = [p_\mu, \hat f] \blacktriangleright 1, \quad \forall \hat f \in \hat{\mathcal A} \\
p_\mu \blacktriangleright 1 = 0
\end{gather}

It follows that
\begin{equation}\begin{split}
\hat{\mathcal H}\blacktriangleright 1 = \hat{\mathcal A}, \\
\hat{\mathcal A}\blacktriangleright 1 = \hat{\mathcal A}.
\end{split}\end{equation}

In a complete analogy, the action $\triangleright$ is a map  $\triangleright:  \mathcal H \otimes \mathcal A \rightarrow \mathcal A$ satisfying the following properties:
\begin{gather}
f \triangleright g = f g, \quad \forall f, g \in \mathcal A\\
p_\mu \triangleright f = [p_\mu, f] \triangleright 1, \quad \forall f \in \mathcal A \\
p_\mu \triangleright 1 = 0
\end{gather}

Also, it follows that
\begin{equation}\begin{split}
\mathcal H \triangleright 1 = \mathcal A, \\
\mathcal A \triangleright 1 = \mathcal A.
\end{split}\end{equation}

$\blacktriangleright$ and $\triangleright$ are actions, so they satisfy:
\begin{gather}
(\hat f \hat g)\blacktriangleright \hat h = \hat f \blacktriangleright (\hat g \blacktriangleright \hat h) \\
(fg)\triangleright h = f \triangleright (g \triangleright h)
\end{gather}

\subsection{Star product}

For $\kappa$-Minkowski space, there exists an isomorphism (as vector spaces) between $\hat{\mathcal A}$ and $\mathcal A$, defined by:
\begin{equation}\begin{split}
\hat f \triangleright 1 &= f \\
f \blacktriangleright 1 &= \hat f,
\end{split}\end{equation}
where $\hat f \in \hat{\mathcal A}$, and also, using realization for $\hat x_\mu$ \eqref{generic-realization}, $\hat f \in \mathcal H$. Similarly, $f \in \mathcal A$, and also inverting realization for $\hat x_\mu$, $f \in \hat{\mathcal H}$.

Using this identification, star~product $\star:\mathcal A \otimes \mathcal A \rightarrow \mathcal A$ is defined by:
\begin{equation}
f \star g \equiv (\hat f \hat g)\triangleright 1 = \hat f \triangleright g
\end{equation}
For $\kappa$-Minkowski space, the star product is associative:
\begin{equation}
(f \star g) \star h = f \star (g \star h).
\end{equation}

Star product defines algebra $\mathcal A_\star$, which is defined like $\mathcal A$, but with non-commutative star product instead of ordinary multiplication. Algebras $\mathcal A_\star$ and $\hat{\mathcal A}$ are isomorphic as algebras, not only as vector spaces. It follows that:
\begin{gather}
(f \star g) \blacktriangleright 1 = \hat f \hat g \\
\begin{split}
\hat{\mathcal H} \triangleright 1 = \mathcal A_\star, \\
\hat{\mathcal A} \triangleright 1 = \mathcal A_\star, 
\end{split}\\
\begin{split}
\mathcal H \blacktriangleright 1 = \hat{\mathcal A}, \\
\mathcal A \blacktriangleright 1 = \hat{\mathcal A}.
\end{split}
\end{gather}


It can be shown that:
\begin{equation}
e^{ik\cdot \hat x} \triangleright 1=e^{iK(k) \cdot x}
\end{equation}
where $K_\mu(k)$ is invertible function $K_\mu: \mathbb M^n \rightarrow \mathbb M^n$, which is calculated in Appendix \ref{AppA} for linear realizations of $\hat x_\mu$. The inverse relation is
\begin{equation}
e^{ik\cdot x} \blacktriangleright 1 =e^{iK^{-1}(k) \cdot \hat x}
\end{equation}

It can also be shown that there is a 
function $P_\mu(k_1,k_2)$, such that
\begin{equation}
e^{ik_1 \cdot \hat x} \triangleright e^{ik_2 \cdot x} = e^{iP(k_1,k_2)\cdot x}
\label{expPkk}
\end{equation}
The star product between such exponentials is then given by:
\begin{equation}
e^{ik_1 \cdot x} \star e^{ik_2 \cdot x} = e^{iK^{-1}(k_1) \cdot \hat x} \triangleright e^{ik_2 \cdot x} = e^{iP(K^{-1}(k_1),k_2) \cdot x} 
\equiv e^{i\mathpzc D(k_1,k_2) \cdot x}
\end{equation}
Note that $K_\mu(k)=P_\mu(k,0)$ and $\mathpzc D_\mu(k_1,k_2) = P_\mu(K^{-1}(k_1),k_2)$. $\mathpzc D_\mu(k_1, k_2)$ describes deformed addition of momenta $(k_1)_\mu \oplus (k_2)_\mu = \mathpzc D_\mu(k_1, k_2)$ (for more details see \cite{ms06}). Calculation of $P_\mu(k_1, k_2)$, $\mathpzc D_\mu(k_1, k_2)$ and $K_\mu(k)$ for linear realizations (described in previous section) is given in Appendix \ref{AppA}.

For elements $f,g\in\mathcal A$ which can be Fourier transformed
\begin{gather}
f=\int \!\mathrm d^nk \ \tilde f(k) e^{i k\cdot x}  \\
g=\int \!\mathrm d^nk \ \tilde g(k) e^{i k\cdot x} 
\end{gather}
we find corresponding elements $\hat f, \hat g \in \hat{\mathcal A}$
\begin{gather}
\hat f = f \blacktriangleright 1 = \int \!\mathrm d^nk \ \tilde f(k) e^{iK^{-1}(k)\cdot\hat x} \\
\hat g = g \blacktriangleright 1 = \int \!\mathrm d^nk \ \tilde g(k) e^{iK^{-1}(k)\cdot\hat x}
\end{gather}
Then the star product $f\star g$ can be written in the following way:
\begin{equation}
f\star g = \hat f \hat g \triangleright 1 = \int \!\mathrm d^nk_1 \mathrm d^nk_2 \ \tilde f(k_1) \tilde g(k_2) e^{i\mathpzc D(k_1,k_2)\cdot x}
\label{starintegral}
\end{equation}

\subsection{Coproduct of momenta}\label{cofmomenta}

The undeformed coproduct $\Delta_0: \mathcal T \rightarrow \mathcal T \otimes \mathcal T$ for momentum $p_\mu$ is:
\begin{equation}
\Delta_0 p_\mu = p_\mu \otimes 1 + 1 \otimes p_\mu \label{Delta0p}
\end{equation}
Deformed coproduct for momenta $\Delta: \mathcal T \rightarrow \mathcal T \otimes \mathcal T$ is \cite{Meljanac09, Meljanac09a, Meljanac2011}:
\begin{equation}
\Delta p_\mu = \mathpzc D_\mu (p \otimes 1, 1 \otimes p)
\end{equation}

Using results from Appendix \ref{AppA}, we have:
\begin{equation}
\Delta p_\mu = p_\mu \otimes 1 + \Lambda^{-1}_{\alpha\mu} \otimes p^\alpha
\label{alldeltap}
\end{equation}
where
\begin{gather}
\Lambda_{\mu\nu}=(e^{\EuScript K})_{\mu\nu}\\
\mathcal K_{\mu\nu} = -K_{\mu\alpha\nu}(K^{-1})^\alpha(p)
\end{gather}
and
\begin{gather}
\Delta \Lambda_{\mu\nu} = \Lambda_ {\mu\alpha} \otimes {\Lambda^\alpha}_\nu \\
\Delta (\Lambda^{-1})_{\mu\nu} = (\Lambda^{-1})_ {\alpha\nu} \otimes {(\Lambda^{-1})_\mu}^\alpha
\end{gather}

We also have:
\begin{gather}
p_\mu \hat f = (p_\mu \blacktriangleright \hat f) + (\Lambda^{-1}_{\alpha\mu} \blacktriangleright \hat f) p^\alpha \\
\Lambda_{\mu\nu} \hat f =  (\Lambda_{\mu\alpha} \blacktriangleright \hat f) {\Lambda^\alpha}_\nu \\
(\Lambda^{-1})_{\mu\nu} \hat f =  ((\Lambda^{-1})_{\alpha\nu} \blacktriangleright \hat f) 
{(\Lambda^{-1})_\mu}^\alpha 
\end{gather}

For example if $\hat f = \hat x_\lambda$ we have
\begin{gather}
[\Lambda_{\mu\nu},\hat x_\lambda]=i{K_{\mu\lambda}}^\alpha \Lambda_{\alpha\nu} \\
[\Lambda^{-1}_{\mu\nu},\hat x_\lambda]=-i\Lambda^{-1}_{\mu\alpha}{K^\alpha}_{\lambda\nu} \label{eq66}
\end{gather}

In order to specify $\Delta p_\mu$, we have to express $K^{-1}_\mu(p)\equiv p^W_\mu$ in terms of momenta $p_\mu$ (see Appendix \ref{AppA}). Momentum $p_\mu$ acts on $e^{ik\cdot x}$ and $e^{ik\cdot \hat x}$ with $\triangleright$ and $\blacktriangleright$ respectively in the following way:
\begin{equation} \label{ppact}
\qquad  p_\mu \triangleright e^{ik\cdot x} = k_\mu e^{ik\cdot x}, \qquad p_\mu \blacktriangleright e^{ik\cdot \hat x} = K_\mu(k) e^{ik\cdot \hat x}
\end{equation}
and momentum $p^W_\mu$ acts as:
\begin{equation} \label{pWact}
\qquad p^W_\mu \blacktriangleright e^{ik\cdot \hat x} = k_\mu e^{ik\cdot \hat x}, \qquad p^W_\mu \triangleright e^{ik\cdot x} = K^{-1}_\mu(k) e^{ik\cdot x}
\end{equation}

It is useful to introduce the shift operator $Z$, with properties:
\begin{gather}
[Z,\hat x_\mu] = ia_\mu Z \\
Z=e^{-a\cdot p^W}
\end{gather}


Explicitly, for $\mathcal C_1$, $\mathcal C_2$, $\mathcal C_3$ and $\mathcal C_4$, coproducts of momenta are:
\begin{itemize}
\item Case $\mathcal C_1$:
\begin{gather}
\label{deltap1}
\Delta p_\mu = p_\mu \otimes 1 + Z \otimes p_\mu + \frac{a_\mu}{a^2}(Z^{1-c}-Z)\otimes a\cdot p \\
\label{lambdainv1}
\Lambda^{-1}_{\mu\nu} = \left[\eta_{\mu\nu} + \frac{a_\mu a_\nu}{a^2}(Z^{-c}-1)\right]Z \\ 
\label{lambda1}
\Lambda_{\mu\nu} = \left[\eta_{\mu\nu} + \frac{a_\mu a_\nu}{a^2}(Z^c-1)\right]Z^{-1} \\
\label{pwc1}
p^W_\mu=\left[p_\mu-\frac{a_\mu}{a^2}(Z-1+a\cdot p)\right] \frac{\ln Z}{Z-1} \\
Z=\left[1-(1-c)a\cdot p \right]^{\frac1{1-c}}
\end{gather}

\item Case $\mathcal C_2$:
\begin{gather}
\label{deltap2}
\Delta p_\mu = p_\mu \otimes 1 + \left( Z^c-\frac c{1+c} \right)\otimes p_\mu
+ \left( c \frac{a_\mu}{a^2} + (c-1)\frac{p^W_\mu}{\ln Z} \right) \frac{Z^{-1}-Z^c}{1+c} \otimes a\cdot p \\
\label{lambdainv2}
\Lambda^{-1}_{\mu\nu}= \eta_{\mu\nu}\left( Z^c-\frac c{1+c} \right) 
  + a_\mu\left( c \frac{a_\nu}{a^2} + (c-1)\frac{p^W_\nu}{\ln Z} \right) \frac{Z^{-1}-Z^c}{1+c} \\
 \label{lambda2}
\Lambda_{\mu\nu}= \eta_{\mu\nu}\left( Z^{-c}-\frac c{1+c} \right) 
  + a_\mu\left( c \frac{a_\nu}{a^2} + (c-1)\frac{p^W_\nu}{\ln Z} \right) \frac{Z-Z^{-c}}{1+c} \\
\label{pwc2}
p^W_\mu=\left[p_\mu-\frac{a_\mu}{a^2}(1-Z^{-1}+a\cdot p)\right] \frac{\ln Z}{1-Z^{-1}} \\
Z=\left[1-(c-1)a\cdot p \right]^{\frac1{c-1}}
\end{gather}

\item Case $\mathcal C_3$:
\begin{gather}
\label{deltap3}
\Delta p_\mu = p_\mu \otimes 1 + Z \otimes p_\mu + a_\mu
\left(
 (1+c)\frac{p^W_\alpha}{\ln Z} -c\frac{a_\alpha}{a^2}
\right)(Z-1)Z \otimes p^\alpha \\
\label{lambdainv3}
\Lambda^{-1}_{\mu\nu}=\left[\eta_{\mu\nu}+\left((1+c)\frac{p^W_\mu}{\ln Z}-c\frac{a_\mu}{a^2}\right)
a_\nu
(Z-1)
\right]Z
\\
\label{lambda3}
\Lambda_{\mu\nu}=\left[\eta_{\mu\nu}+\left((1+c)\frac{p^W_\mu}{\ln Z}-c\frac{a_\mu}{a^2}\right)a_\nu
(Z^{-1}-1)
\right]Z^{-1} \\
\label{pwc3}
p^W_\mu=\left[p_\mu-\frac{a_\mu}{a^2}(Z-1+a\cdot p)\right] \frac{\ln Z}{Z-1} \\
Z=\left[ c + (1-c)\left( (1-a\cdot p)^2-a^2 p^2 \right) \right]^{\frac1{2(1-c)}}
\end{gather}

\item Case $\mathcal C_4$:
\begin{gather}
\label{deltap4}
\Delta p_\mu = p_\mu \otimes 1 + 1 \otimes p_\mu + p_\mu \otimes a\cdot p -a_\mu p_\alpha Z \otimes p^\alpha - \frac{a_\mu}2p^2Z\otimes a\cdot p \\
\label{lambdainv4}
\Lambda^{-1}_{\mu\nu}=\eta_{\mu\nu} + a_\mu p_\nu - \left(p_\mu + \frac{a_\mu}2p^2 \right)a_\nu Z \\
\label{lambda4} 
\Lambda_{\mu\nu}=\eta_{\mu\nu} + p_\mu a_\nu - a_\mu\left(p_\nu + \frac{a_\nu}2p^2 \right) Z \\
\label{pwc4}
p^W_\mu=\left(p_\mu +\frac{a_\mu}2p^2 \right) \frac{\ln Z}{1-Z^{-1}} \\
Z=\frac1{1+a\cdot p}
\end{gather}

\end{itemize}

\subsection{Relation between star product, twist operator and coproduct}

The star product is related to the twist operator $\mathcal F^{-1}$ in the following way:
\begin{equation}
f \star g = m \left[ \mathcal F^{-1} (\triangleright \otimes \triangleright)(f \otimes g) \right]
\label{startwist}
\end{equation}
where $f,g \in \mathcal A$. Furthermore,
\begin{equation}
\hat f = m \left[ \mathcal F^{-1} (\triangleright \otimes 1)(f \otimes 1) \right], \quad f \in \mathcal A
\end{equation}
where $\hat f \in \hat{\mathcal A}$ is expressed in terms of $x, p \in \mathcal H$.

Using the above expression for star product eq. \eqref{starintegral} and \eqref{startwist}, the twist operator can be written as \cite{ms06, Govindarajan-2, govindarajan}
\begin{equation}
\mathcal F^{-1} = :\exp\left[i(tx_\alpha\otimes1 + (1-t)\otimes x_\alpha)(\Delta-\Delta_0)p^\alpha \right]:
\label{normaltwist}
\end{equation}
where $t\in\mathbb R$ and is generally defined up to the right ideal $\mathcal I_0 \subset \mathcal H \otimes \mathcal H$ defined by
\begin{equation}
m\left( \mathcal I_0 (\triangleright \otimes \triangleright)(\mathcal A \otimes \mathcal A) \right) = 0
\end{equation}

\section{Drinfeld twists}

Starting with expression \eqref{normaltwist} for twist operator, we derive Drinfeld twists \cite{8, Drinfeld, Drinfeld1} in Appendix \ref{AppB}.
\begin{equation}
\mathcal F= \exp\left( \mathcal K_{\beta\alpha} \otimes L^{\alpha\beta} \right) = \exp\left( -ip^W_\alpha \otimes \kl^\alpha \right)
\label{twistF}
\end{equation}
where $p^W_\mu$ is given in the subsection \ref{cofmomenta} after equation \eqref{eq66} and in Appendix A, and $\kl_\mu=-iK_{\beta\mu\alpha}L^{\alpha\beta}$, where $K_{\beta\mu\alpha}$ satisfies \eqref{KisC} and \eqref{KKCK}, with solutions eq. \eqref{cetri}, $\kl_\mu$ generate $\kappa$-Minkowski algebra
\begin{equation}
[\kl_\mu, \kl_\nu] = i(a_\mu\kl_\nu - a_\nu\kl_\mu)
\end{equation}
and $L_{\mu\nu}$ generate $\mathfrak{gl}(n)$ algebra, see equation \eqref{igln}.

The classical r-matrix $r_{cl}$, related to twist \eqref{twistF} is:
\begin{equation}
r_{cl}= p_\alpha \wedge l^\alpha = p_\alpha \otimes l^\alpha - l^\alpha \otimes p_\alpha 
\end{equation}
For $\mathcal C_1$, $\mathcal C_2$, $\mathcal C_3$ and $\mathcal C_4$, twists are: 
\begin{align}
\label{twistC1}
&\mathcal F_{\mathcal C_1} = \exp \left\{
-\left( \eta_{\alpha\beta} - c\frac{a_\alpha a_\beta}{a^2} \right) \ln Z
\otimes L^{\alpha\beta} \right\}
=\exp \left\{
-\ln Z \otimes \left(D-c\frac{a_\alpha a_\beta}{a^2}L^{\alpha\beta} \right)
\right\} 
, \\
\label{twistC2}
&\mathcal F_{\mathcal C_2} = \exp \left\{
-\left[
c\left( \eta_{\alpha\beta} - \frac{a_\alpha a_\beta}{a^2} \right) \ln Z
+(1-c)a_\beta p^W_\alpha
\right]
\otimes L^{\alpha\beta} \right\}, \\
\label{twistC3}
&\mathcal F_{\mathcal C_3} = \exp \left\{
-\left[
\left( \eta_{\alpha\beta} - c\frac{a_\alpha a_\beta}{a^2} \right) \ln Z
- (1+c)a_\alpha p^W_\beta
\right]
\otimes L^{\alpha\beta} \right\}, \\
\label{twistC4}
&\mathcal F_{\mathcal C_4} =
\exp \left\{
\left( a_\alpha p^W_\beta-a_\beta p^W_\alpha \right) \otimes L^{\alpha\beta}
\right\} = 
\exp \left\{
a_\alpha p^W_\beta \otimes M^{\alpha\beta}
\right\}, \quad a^2=0
\end{align}
Note that only for the case $\mathcal C_4$ ($a^2=0$), the corresponding twist operator can be expressed in terms of Poincar\'e generators only \cite{universal, forme}.

Starting from the twist operator, the realization can be obtained using
\begin{equation}
\hat x_\mu = m\left[\mathcal F^{-1} (\triangleright \otimes 1)(x_\mu \otimes 1) \right] =
 x_\mu - iK_{\beta\mu\alpha}L^{\alpha\beta}
\end{equation}
Using twists \eqref{twistC1}, \eqref{twistC2}, \eqref{twistC3} and \eqref{twistC4} yields realizations $\mathcal C_1$, $\mathcal C_2$, $\mathcal C_3$ and $\mathcal C_4$ respectively, which satisfy $\kappa$-Minkowski algebra.

\subsection{Undeformed $\mathfrak{igl}(n)$ Hopf algebra}\label{UndefHA}
Coproducts $\Delta_0: \mathfrak{igl}(n) \rightarrow \mathfrak{igl}(n) \otimes \mathfrak{igl}(n)$ in undeformed $\mathfrak{igl}(n)$ Hopf algebra are:
\begin{gather}
\Delta_0 p_\mu = p_\mu \otimes 1 + 1 \otimes p_\mu \\
\Delta_0 L_{\mu\nu} = L_{\mu\nu} \otimes 1 + 1 \otimes L_{\mu\nu},
\end{gather}
counit $\epsilon: \mathfrak{igl}(n)\rightarrow \mathbb C$ is
\begin{equation}
\epsilon(p_\mu)=\epsilon(p^W_\mu)=\epsilon(L_{\mu\nu})=0, \qquad \epsilon(1)=1
\end{equation}
and antipode $S_0: \mathfrak{igl}(n) \rightarrow \mathfrak{igl}(n)$ is
\begin{equation}
S_0(p_\mu)=-p_\mu, \qquad S_0(L_{\mu\nu})=-L_{\mu\nu}.
\end{equation}

\subsection{Normalization condition}
Now we show that these twists satisfy normalization condition and cocycle condition, i.e. that they are Drinfeld twists.

The normalization condition 
\begin{equation}
m(\epsilon\otimes1)\mathcal F = 1 = m(1\otimes\epsilon)\mathcal F
\end{equation}
follows trivially since the twist is of the form $\mathcal F=e^f$, where $f= -ip^W_\alpha \otimes \kl^\alpha$, therefore,
\begin{equation}
(\epsilon\otimes1)f=(1\otimes\epsilon)f=0,
\end{equation}
and from this follows
\begin{gather}
(\epsilon\otimes1)\mathcal F = (\epsilon\otimes1)e^f = 1\otimes1, \\
(1\otimes\epsilon)\mathcal F = (1\otimes\epsilon)e^f = 1\otimes1.
\end{gather}

\subsection{Cocycle condition}

Cocycle condition is
\begin{equation}
(\mathcal F \otimes 1)(\Delta_0 \otimes 1)\mathcal F = (1\otimes\mathcal F)(1\otimes\Delta_0)\mathcal F
\label{cocycle}
\end{equation}

We shall prove it using factorization properties of twist $\mathcal F$
\begin{gather}
(\Delta \otimes 1)\mathcal F = \mathcal F_{23} \mathcal F_{13} \label{1ff} \\
(1\otimes\Delta_0)\mathcal F = \mathcal F_{13} \mathcal F_{12} \label{2ff}
\end{gather}
where
\begin{gather}
\mathcal F_{12} = e^{\mathcal K_{\beta\alpha} \otimes L^{\alpha\beta} \otimes 1} \\
\mathcal F_{13} = e^{\mathcal K_{\beta\alpha} \otimes 1 \otimes L^{\alpha\beta}} \\
\mathcal F_{23} = e^{1 \otimes \mathcal K_{\beta\alpha} \otimes L^{\alpha\beta}}.
\end{gather}

The first factorization property \eqref{1ff} can be proven to hold in a following way:
\begin{equation}\begin{split}
\mathcal F_{23} \mathcal F_{13} 
&= e^{1\otimes (-ip^W_\alpha) \otimes \kl^\alpha}e^{-ip^W_\alpha \otimes 1 \otimes \kl^\alpha} \\
&= \left( e^{ip^W_\alpha \otimes 1 \otimes \kl^\alpha} e^{1\otimes ip^W_\alpha \otimes \kl^\alpha} \right)^{-1} \\
&=e^{-i\mathpzc D^W_\alpha(p^W \otimes 1 , 1\otimes p^W) \otimes \kl^\alpha} 
\end{split}\end{equation}
This holds because $\kl_\mu$ generates the same algebra as $\hat x_\mu$ and 
\begin{equation}
e^{ik_1\cdot\hat x}e^{ik_2\cdot\hat x}=e^{i\mathpzc D^W(k_1,k_2)\cdot\hat x}.
\end{equation}
Furthermore, since $\mathpzc D^W_\mu(p^W \otimes 1 , 1\otimes p^W)=\Delta p^W_\mu$, it follows that:
\begin{equation}
\mathcal F_{23} \mathcal F_{13} =e^{-i\Delta p^W_\alpha \otimes \kl^\alpha} 
=(\Delta \otimes 1)e^{-ip^W_\alpha\otimes \kl^\alpha} 
=(\Delta \otimes 1)\mathcal F
\end{equation}

The second factorization property \eqref{2ff} for our twist follows trivially:
\begin{equation}\begin{split}
(1\otimes\Delta_0)\mathcal F &= (1\otimes\Delta_0)e^{\mathcal K^{\beta\alpha}\otimes L_{\alpha\beta}} \\
&=e^{\mathcal K^{\beta\alpha}\otimes (\Delta_0 L_{\alpha\beta})} \\
&=e^{\mathcal K^{\beta\alpha}\otimes 1 \otimes L_{\alpha\beta}}
e^{\mathcal K^{\beta\alpha} \otimes L_{\alpha\beta}\otimes 1}\\
&=\mathcal F_{13} \mathcal F_{12}
\end{split}\end{equation}


To see that cocycle condition follows from factorization properties, the first property, eq. \eqref{1ff}, should be multiplied by $\mathcal F_{12}$ from the right and second one, eq. \eqref{2ff}, by $\mathcal F_{23}$ from the left:
\begin{gather}
[(\Delta \otimes 1)\mathcal F] (\mathcal F \otimes 1) = \mathcal F_{23} \mathcal F_{13} \mathcal F_{12} \\
(1\otimes\mathcal F)(1\otimes\Delta_0)\mathcal F = \mathcal F_{23} \mathcal F_{13} \mathcal F_{12}
\end{gather}
which implies
\begin{equation}
[(\Delta \otimes 1)\mathcal F] (\mathcal F \otimes 1) = (1\otimes\mathcal F)(1\otimes\Delta_0)\mathcal F
\end{equation}

Since $[(\Delta \otimes 1)\mathcal F] (\mathcal F \otimes 1)
= (\mathcal F \otimes 1)(\Delta_0 \otimes 1)\mathcal F$, this is the cocycle condition \eqref{cocycle}.

\subsection{R-matrix}
R-matrix is defined by \cite{Govindarajan-2, rina}:
\begin{equation}
\mathcal R=\tilde {\mathcal F} \mathcal F^{-1}=
e^{-i \kl^\alpha \otimes p^W_\alpha}e^{ip^W_\beta \otimes \kl^\beta}=1\otimes 1+r_{cl}+\mathcal O\left(\frac{1}{\kappa^2}\right)
\end{equation}
where $\tilde{\mathcal F}=\tau_0\mathcal F\tau_0$ is a transposed twist, see section VII. for details.

Up to the second order we have:
\begin{equation}
\ln\mathcal{R}=i(p^W_\alpha \otimes \kl^\alpha - \kl^\alpha \otimes p^W_\alpha)
-\frac12\left(
[p^W_\alpha, \kl^\beta] \otimes \kl^\alpha p^W_\beta -
\kl^\alpha p^W_\beta \otimes [p^W_\alpha, \kl^\beta]
\right) +\mathcal O(K^3)
\end{equation}
where
\begin{equation}
[p^W_\mu, \kl_\nu]=[p^W_\mu, \hat x_\nu-x_\nu]=[p^W_\mu, \hat x^\alpha](\eta_{\alpha\nu}-\varphi^{-1}_{\alpha\nu}(p))
\end{equation}
Commutator $[p^W_\mu, \hat x_\nu]$ is given in equation \eqref{pwhatx} and inverse matrices $\varphi^{-1}_{\mu\nu}$ are given in \eqref{inversephi} and relation between $p_\mu$ and $p^W_\mu$ is given in equation \eqref{ppW}.

Generally, classical matrix $r_{cl}=\ln\mathcal{R}$ up to the first order in $\frac{1}{\kappa}$ and the classical $r_{cl}$ matrices can be written in terms of $\mathfrak{igl}(n)$ generators as
\begin{equation}\label{rcl}
r_{cl}=p_{\mu}\wedge l^{\mu}=-iK_{\beta\mu\alpha}p^{\mu}\wedge L^{\alpha\beta},
\end{equation}
where $K_{\beta\mu\alpha}$ 
are given in \eqref{cetri}.
Using \eqref{rcl} we find the classical $r_{cl}$-matrices for twists \eqref{twistC1}, \eqref{twistC2}, \eqref{twistC3} and \eqref{twistC4}:
\begin{align}
r_{cl}^{(\mathcal C_1)} & = a\cdot p \wedge \left[\left(1-\frac cn\right)D - c \frac{a_\alpha a_\beta}{a^2}S^{\alpha\beta} \right]\\
r_{cl}^{(\mathcal C_2)} & = a\cdot p \wedge \left[ \left(c-\frac1n\right)D-c\frac{a^\alpha a^\beta}{a^2}S_{\alpha\beta} \right]
-(1-c)p^\alpha \wedge a^\beta \left(S_{\alpha\beta}+\frac12M_{\alpha\beta}\right)\\
r_{cl}^{(\mathcal C_3)} & = a\cdot p \wedge \left[\left(1+\frac1n\right)D - c \frac{a_\alpha a_\beta}{a^2}S^{\alpha\beta} \right] 
+(1+c)p^\alpha \wedge a^\beta \left(S_{\alpha\beta}-\frac12M_{\alpha\beta}\right)\\
r_{cl}^{(\mathcal C_4)} & = a_\alpha P_\beta \wedge M^{\alpha\beta} \label{rcl4}
\end{align}
where 
\begin{equation}
S_{\mu\nu} = \frac12(L_{\mu\nu}+L_{\nu\mu})-\frac1nD\eta_{\mu\nu}
\end{equation}
is  traceless symmetric part of $L_{\mu\nu}$, $D=L^{\alpha}_{\ \alpha}$ and $M_{\mu\nu}=L_{\mu\nu}-L_{\nu\mu}$. 
Note that for the case $\mathcal{C}_4$, $r^{(\mathcal{C}_4)}_{cl}$ in \eqref{rcl4} coincides with the $r_{cl}$ for light-cone case discussed in \cite{BorPach}. Also $r_{cl}^{(\mathcal C_1)}=r_{cl}^{(\mathcal C_2)}$ for $c=1$ and $r_{cl}^{(\mathcal C_1)}=r_{cl}^{(\mathcal C_3)}$ for $c=-1$, which is consistent with discussion in section III.


\section{Twisted symmetry algebras}

The family of twists \eqref{twistF}, applied to undeformed $\mathfrak{igl}(n)$ Hopf algebra (subsection \ref{UndefHA}) produces the corresponding $\kappa$-deformed $\mathfrak{igl}(n)$ Hopf algebras. 
For $h \in \mathfrak{igl}(n)$, deformed coproduct $\Delta h$ is related to undeformed coproduct $\Delta_0 h$ via:
\begin{equation}
\Delta h = \mathcal F \Delta_0 h \mathcal F^{-1}
\end{equation}

In deformed $\mathfrak{igl}(n)$ Hopf algebra 
the coproduct $\Delta$ is:
\begin{align} 
\label{Deltap}
\Delta p_\mu &=  
 \mathcal F \Delta_0 p_\mu \mathcal F^{-1} =
p_\mu \otimes 1 + \Lambda^{-1}_{\alpha\mu} \otimes p^\alpha \\ \label{DeltaL}
\Delta L_{\mu\nu}&=
 \mathcal F \Delta_0 L_{\mu\nu} \mathcal F^{-1}=
L_{\mu\nu}\otimes1+\left(\Lambda^{-1}_{\beta\gamma} \frac{\partial \Lambda^\gamma{}_\alpha}{\partial p^\mu}p_\nu+\Lambda^{-1}_{\beta\nu}\Lambda_{\mu \alpha}\right)\otimes L^{\alpha\beta} \\
\Delta \Lambda_{\mu\nu} &= \Lambda_{\mu\alpha}\otimes {\Lambda^\alpha}_\nu \\
\Delta (\Lambda^{-1})_{\mu\nu} &= (\Lambda^{-1})_ {\alpha\nu} \otimes {(\Lambda^{-1})_\mu}^\alpha
\end{align}
where $\Lambda_{\mu\nu}$ and $\Lambda^{-1}_{\mu\nu}$ are given in equations \eqref{lambda1}, \eqref{lambdainv1}, \eqref{lambda2}, \eqref{lambdainv2}, \eqref{lambda3}, \eqref{lambdainv3}, \eqref{lambda4} and \eqref{lambdainv4} for $\mathcal C_1$, $\mathcal C_2$, $\mathcal C_3$ and $\mathcal C_4$ respectively. 

We point out that generators $\kl_\mu$ (see equation \eqref{mKxp}) close $\kappa$-Minkowski algebra (see equation \eqref{commm}) and $[\kl_\mu,p_\nu]=iK_{\alpha\mu\nu}p^\alpha$. Note that twists can be expressed in terms of $\kl_\mu$ and $p_\mu$. From this it follows that $\Delta \kl_\mu=\mathcal F \Delta_0 \kl_\mu \mathcal F^{-1}$ (where $\Delta_0 \kl_\mu = \kl_\mu \otimes 1+1\otimes \kl_\mu$) is closed in $\kl_\mu$ and $p_\mu$.

The counit is unchanged:
\begin{equation}
\epsilon(p_\mu)=\epsilon(L_{\mu\nu})=0, \qquad \epsilon(\Lambda_{\mu\nu})=\epsilon(\Lambda^{-1}_{\mu\nu})=\eta_{\mu\nu}
\end{equation}
and the antipode $S$, obtained from coproduct and counit via $m[(S\otimes1)\Delta h]=m[(1\otimes S)\Delta h]=\epsilon(h)$, is given by
\begin{align}
\label{Sp}
S(p_\mu)&=  -\Lambda^{-1}_{\alpha\mu} p^\alpha \\
\label{SL}
S(L_{\mu\nu})&= -\left(
{\Lambda_\beta}^\gamma \frac{\partial \Lambda^{-1}_{\gamma\alpha}}{\partial S(p^\mu)}S(p_\nu)
+\Lambda_{\beta\nu}\Lambda^{-1}_{\mu \alpha}
\right)L^{\alpha\beta} \\
S(\Lambda_{\mu\nu})&= \Lambda^{-1}_{\mu\nu} \\
S(\Lambda^{-1}_{\mu\nu})&= \Lambda_{\mu\nu}
\end{align}


The deformed Hopf algebra acting on $\hat x_\mu \otimes 1$, i.e. using $g\hat f = m\left[\Delta g(\blacktriangleright \otimes 1)(\hat f \otimes 1) \right]$, $\forall g \in \mathfrak{igl}(n)$ and $\hat f \in \hat{\mathcal A}$ leads to
\begin{align}
[L_{\rho\sigma}, \hat{x}_{\nu}]&=\eta_{\sigma\nu}\hat{x}_{\rho}+i\eta_{\sigma\nu}K_{\mu\rho\alpha}L^{\alpha\mu}-iK_{\mu\nu\sigma}L_{\rho}^{\ \mu}+iK_{\rho\nu\alpha}L^{\alpha}_{\ \sigma} \\
[p_{\mu}, \hat{x}_{\nu}]&=-i(\eta_{\mu\nu}+K_{\beta\nu\mu}p^{\beta})
\end{align}
which also leads to \eqref{commx}.

Let us consider special cases. For the case $\mathcal S_1$, the twist operator is: 
\begin{equation} \label{twistS1}
\mathcal F_{\mathcal S_1} =\exp \left\{ -\ln(1-a\cdot p)\otimes D \right\}
\end{equation}
and coproducts and antipodes of $p_\mu$, $D\equiv {L_\alpha}^\alpha$ and $M_{\mu\nu}$, obtained from the twist \eqref{twistS1}, are:
\begin{equation}\begin{split} \label{S1alg}
\Delta p_\mu &= p_\mu \otimes 1 + Z \otimes p_\mu = \Delta_0 p_\mu - a\cdot p \otimes p_\mu \\
\Delta D &= D\otimes 1 + Z^{-1}\otimes D \\
\Delta M_{\mu\nu}&=\Delta_0 M_{\mu\nu} + (a_\mu p_\nu - a_\nu p_\mu)Z^{-1}\otimes D \\
S(p_\mu)&=-Z^{-1}p_\mu \\
S(D)&=-ZD = -D + (a\cdot p) D \\
S(M_{\mu\nu})&=-M_{\mu\nu}+(a_\mu p_\nu - a_\nu p_\mu)
\end{split}\end{equation}
The coproduct and antipode of $\kl_\mu$ are:
\begin{equation}\begin{split}
\Delta \kl_\mu&=\kl_\mu \otimes 1 + Z^{-1}\otimes\kl_\mu \\
S(\kl_\mu)&=-Z\kl_\mu
\end{split}\end{equation}
The symmetry of this case is Poincar\'e-Weyl symmetry. The case $\mathcal S_1$ corresponds to the right covariant realization $\hat x_\mu=x_\mu-a_\mu(x\cdot p)$, see equation \eqref{specialxp}, and is related to \cite{22}, but with interchanged left and right side in tensor product and with $a_\mu \rightarrow -a_\mu$.

For the case $\mathcal S_2$, the twist operator is: 
\begin{equation} \label{twistS2}
\mathcal F_{\mathcal S_2} = \exp \left\{ -a_\beta p_\alpha \frac{\ln(1+a\cdot p)}{a\cdot p} \otimes L^{\alpha\beta} \right\} 
\end{equation}
and coproducts and antipodes of $p_\mu$ and $L_{\mu\nu}$ are:
\begin{equation}\begin{split} \label{S2alg}
\Delta p_\mu &= \Delta_0 p_\mu + p_\mu\otimes a\cdot p = p_\mu\otimes Z^{-1}+1\otimes p_\mu \\ 
\Delta L_{\mu\nu}&= \Delta_0 L_{\mu\nu} - a_\mu p^\alpha Z \otimes L_{\alpha\nu}\\
S(p_\mu)&=-Zp_\mu \\
S(L_{\mu\nu})&= -L_{\mu\nu}-a_\mu p^\alpha L_{\alpha\nu} 
\end{split}\end{equation}
The coproduct and antipode of $\kl_\mu$ are:
\begin{equation}\begin{split}
\Delta \kl_\mu&=\Delta_0\kl_\mu  + a_\mu p_\alpha Z\otimes\kl^\alpha \\
S(\kl_\mu)&=-\kl_\mu-a_\mu(p\cdot\kl)
\end{split}\end{equation}
The case $\mathcal S_2$ corresponds to the left covariant realization $\hat x_\mu=x_\mu\left[1+(a\cdot p)\right]$, see equation \eqref{specialxp}.

For the case $\mathcal S_3$, the twist operator is: 
\begin{equation} \label{twistS3}
\mathcal F_{\mathcal S_3} =\exp \left\{ -\ln Z\otimes D+ a_\alpha p^W_\beta \otimes L^{\alpha\beta} \right\}  
\end{equation}
where
\begin{equation}
p^W_\mu=\left(
p_\mu + \frac{a_\mu p^2}{Z+1-a\cdot p}
\right) \frac{\ln Z}{Z-1}
\end{equation}
and
\begin{equation}
Z=\sqrt{(1-a\cdot p)^2-a^2 p^2}
\end{equation}
and coproduct and antipode of $p_\mu$ are:
\begin{equation}\begin{split} \label{S3alg}
\Delta p_\mu &=  p_\mu \otimes 1 + 
\left[
\eta_{\mu\alpha} + a_\mu \left(p_\alpha - \frac{a_\alpha p^2}{Z-1+a\cdot p} \right)
\right]Z\otimes p^\alpha\\
S(p_\mu)&= -\left(
p_\mu + a_\mu p^2 \frac{Z-1}{Z-1+a\cdot p}
\right)Z
\end{split}\end{equation}
Similarly one finds $\Delta L_{\mu\nu}$ and $S(L_{\mu\nu})$ using equations \eqref{DeltaL} and \eqref{SL} respectively. The coproduct and antipode of $\kl_\mu$ are:
\begin{equation}\begin{split}
\Delta \kl_\mu&=\kl_\mu\otimes 1 +Z\otimes\kl_\mu
+\left[\frac{Z-1}{\ln Z}a_\mu p^W_\alpha+ \frac{a_\alpha p_\mu}{Z^2} \right]\otimes\kl^\alpha \\
S(\kl_\mu)&=-Z^{-1}\kl_\mu + \frac{1-Z^{-1}}{\ln Z}a_\mu(p^W\cdot\kl)+
\left(
p_\mu + a_\mu p^2 \frac{Z-1}{Z-1+a\cdot p}
\right)Z^3(a\cdot\kl)
\end{split}\end{equation}
The case $\mathcal S_3$ corresponds to $\hat x_\mu=x_\mu-a_\mu(x\cdot p)-(a\cdot x)p_\mu$, see equation \eqref{specialxp}.

For the case $\mathcal C_4$, i.e. for the light-like $\kappa$ deformation of Poincar\'e Hopf algebra, the twist operator is: 
\begin{equation} \label{twistC4P}
\mathcal F_{\mathcal C_4} =\exp \left\{ a_\alpha p_\beta \frac{\ln(1+a\cdot p)}{a\cdot p} \otimes M^{\alpha\beta} \right\}
\end{equation}
and coproducts and antipodes of $p_\mu$ and $M_{\mu\nu}$, obtained from the twist \eqref{twistC4P}, are:
\begin{equation}\begin{split} \label{kphalgebra}
\Delta p_\mu &= \Delta_0 p_\mu + \left[
p_\mu a^\alpha - a_\mu
\left( p^\alpha + \frac{1}{2}a^\alpha p^2 \right)Z
\right]\otimes p_\alpha
\\
\Delta M_{\mu\nu} &= \Delta_0 M_{\mu\nu} +
(\delta^\alpha_\mu a_\nu-\delta^\alpha_\nu a_\mu)\left(
p^\beta+\frac{1}{2}a^\beta p^2
\right)Z\otimes M_{\alpha\beta} 
\\
S(p_\mu) &= \left[-p_\mu -a_\mu \left(p_\alpha + \frac{1}{2}a_\alpha p^2 \right) p^\alpha \right]Z \\
S(M_{\mu\nu}) &= 
-M_{\mu\nu} +(-a_\mu \delta^\beta_\nu+a_\nu \delta^\beta_\mu) \left(p^\alpha + \frac{1}{2}a^\alpha p^2 \right)M_{\alpha\beta} 
\end{split}\end{equation}
The coproduct and antipode of $\kl_\mu$ are:
\begin{equation}\begin{split}
\Delta \kl_\mu&=\Delta_0\kl_\mu +a_\mu p_\alpha \otimes\kl^\alpha \\
S(\kl_\mu)&=-\kl_\mu + a_\mu Z (p\cdot\kl)
\end{split}\end{equation}
The case $\mathcal C_4$ corresponds to the natural realization $\hat x_\mu=x_\mu\left[1+(a\cdot p)\right]-(a\cdot x)p_\mu$, see equation \eqref{realizations}. It is the only solution compatible with $\kappa$-Poincar\'e Hopf algebra \cite{BorPach, universal, kovacevic-meljanac, BP2010}. 

\section{Transposed Drinfeld twists and left-right dual $\kappa$-Minkowski algebra}
\label{dualssect}

Transposed twist is $\tilde{\mathcal F}=\tau_0 \mathcal F \tau_0$, where $\tau_0: \mathcal H \otimes \mathcal H \rightarrow \mathcal H \otimes \mathcal H$ is a linear map such that $\tau_0(A\otimes B)=B\otimes A$ $\forall A, B \in \mathcal H$. It is obtained from $\mathcal F$ by interchanging left and right side of tensor product, and it is also a Drinfeld twist satisfying normalization and cocycle condition. It is obtained from \eqref{normaltwist} by taking $t=1$ and using transposed coproduct $\tilde\Delta p_\mu = \tau_0 \Delta p_\mu \tau_0$ instead of $\Delta p_\mu$.

From transposed Drinfeld twist, a set of left-right dual generators of $\kappa$-Minkowski spacetime can be obtained:
\begin{equation} \label{formula7A}
\hat y_\mu = m\left[\tilde{\mathcal F}^{-1} (\triangleright \otimes 1)(x_\mu \otimes 1) \right] =
 x^\alpha\Lambda^{-1}_{\mu\alpha}
\end{equation}
where $\Lambda^{-1}_{\mu\nu}$ for $\mathcal C_1$, $\mathcal C_2$, $\mathcal C_3$ and $\mathcal C_4$ is given in \eqref{lambdainv1}, \eqref{lambdainv2}, \eqref{lambdainv3} and \eqref{lambdainv4} respectively. For example, for cases $\mathcal S_1$, $\mathcal S_2$, $\mathcal S_3$ and $\mathcal C_4$, generators $\hat y_\mu$ and $\hat x_\mu$ are:
\begin{align}
\mathcal S_1\text:& ~ \hat y_\mu = x_\mu(1-a\cdot p),
&\hat x_\mu&=x_\mu-a_\mu(x\cdot p), \\
\mathcal S_2\text:& ~ \hat y_\mu = x_\mu - (a\cdot x)p_\mu,
&\hat x_\mu&= x_\mu(1+a\cdot p), \\
\mathcal S_3\text:& ~ \hat y_\mu = \left[ x_\mu + 
a_\mu \left(
x \cdot p - \frac{(a\cdot x) p^2}{Z+1-a\cdot p}
\right) \right]Z,
&\hat x_\mu&=x_\mu - (a\cdot x)p - a_\mu(x\cdot p), \\
\mathcal C_4\text:& ~ \hat y_\mu = x_\mu + (a\cdot x)p_\mu - a_\mu\left(x\cdot p +\frac{a\cdot x}2 p^2 \right)Z,
&\hat x_\mu&=x_\mu - (a\cdot x)p_\mu - x_\mu (a\cdot p), ~ a^2=0
\end{align}
Generators $\hat y_\mu$ satisfy $\kappa$-Minkowski algebra but with $-a_\mu$ instead of $a_\mu$ \cite{kovacevic-meljanac}:
\begin{equation}
[\hat y_\mu, \hat y_\nu] = -i(a_\mu\hat y_\nu - a_\nu\hat y_\mu)
\end{equation}
Generators of $\kappa$-Minkowski space $\hat x_\mu$ commute with their duals $\hat y_\mu$ \cite{kovacevic-meljanac}:
\begin{equation}
[\hat x_\mu, \hat y_\nu] = 0
\end{equation}

Generally, dual basis $\hat y_\mu$ is related to basis $\hat x_\mu$ via
\begin{equation}
\hat y_\mu=\hat x^\alpha \left( e^{-\mathcal C} \right)_{\mu\alpha}
\end{equation}
where $\mathcal C_{\mu\nu}=-C_{\mu\alpha\nu}(p^W)^\alpha$, where $C_{\mu\alpha\nu}$ are structure constants (see Appendix A in\cite{forme}). From this relation, and equations \eqref{generic-realization} and \eqref{formula7A} for $\hat x_\mu$ and $\hat y_\mu$ respectively, it follows:
\begin{equation}
\Lambda^{-1}_{\mu\nu}=\left( e^{-\mathcal K} \right)_{\mu\nu}=   \left( e^{-\mathcal C} \right)_{\mu\alpha} {\varphi_\nu}^\alpha
\end{equation}

\subsection{$\kappa$-Minkowski algebra from transposed twists with $a_\mu \rightarrow -a_\mu$}
Starting with the family of twists \eqref{twistF}, we define related Drinfeld twists $\left.\tilde{\mathcal F}\right\vert_{a_\mu\rightarrow-a_\mu}$. They lead to nonlinear realizations of $\hat x_\mu$, satisfying \eqref{commx}:
\begin{equation}
\hat x_\mu = m\left[
\left.\tilde{\mathcal F}^{-1}\right\vert_{a_\mu\rightarrow-a_\mu} (\triangleright\otimes1)(x_\mu\otimes1)
\right]
= x^\alpha \left. \Lambda^{-1}_{\mu\alpha}\right\vert_{a_\mu\rightarrow-a_\mu}
\end{equation}
Then the corresponding dual generators $\hat y_\mu$ are given by
\begin{equation}
\hat y_\mu = x_\mu + iK_{\beta\mu\alpha}L^{\alpha\beta} = x_\mu - l_\mu
\end{equation}

Compared to case with transposed twists $\tilde{\mathcal F}$ in the beginning of this section, here the roles of $\hat x_\mu$ and $\hat y_\mu$ are interchanged, with $a_\mu \rightarrow -a_\mu$. With this new family of twists, $\hat x_\mu$ are non-linear realizations of $\kappa$-Minkowski space, while $\hat y_\mu$ are linear realizations of dual $\kappa$-Minkowski space.

If we apply twists $\left.\tilde{\mathcal F}\right\vert_{a_\mu\rightarrow-a_\mu}$ to undeformed coproducts $\Delta_0h$, we get coproducts $\tilde\Delta\vert_{a_\mu\rightarrow-a_\mu}h$, i.e. left and right side in coproducts $\Delta h$ are interchanged and $a_\mu$ is replaced by $-a_\mu$.

We point out that solution $\mathcal C_4$, and its transposed case, are of special interest because they lead to light-like $\kappa$-Poincar\'e Hopf algebra. They are related to the result by 
Borowiec and Pacho\l{} \cite{BorPach} (comparison is given is section \ref{LCdef}). 

\section{Non-linear realizations of $\kappa$-Minkowski space and related Drinfeld twists}

We shall also present a few families of non-linear realizations and corresponding Drinfeld twist operators known in the literature so far. 

\subsection{Timelike deformations}
The realizations we are considering are \cite{ms06, Meljanac-4}:
\begin{align}\hat x_i &= x_i \varphi(A) \\ \hat x_0&= x_0 \psi(A) - a_0 x_k p^k \gamma(A) \end{align}
where $A=-a\cdot p$ and functions $\varphi(A)$, $\psi(A)$ are 
such that 
$\varphi(0)=\psi(0)=1$ and related to $\gamma(A)$ by:
\begin{equation}\label{gammaA}
\gamma(A)=\frac{\psi(A)}{\varphi(A)}\frac{d\varphi(A)}{d A}+1
\end{equation}
Generically, the symmetry algebra is $\kappa$-deformed $\mathfrak{igl}(n)$ Hopf algebra. We will present two cases.


\textbf{i)} The first case is $\psi(A)=1$, with arbitrary $\varphi(A)$ and $\gamma(A)=\frac{\varphi'(A)}{\varphi(A)}+1$, see equation \eqref{gammaA}. The coproducts of momenta are:
\begin{align}
\Delta p_0 &= \Delta_0 p_0 = p_0 \otimes 1 + 1 \otimes p_0 \\
\Delta p_i &= \varphi(A\otimes1+1\otimes A)
\left(
\frac{p_i}{\varphi(A)}\otimes1 + e^A \otimes \frac{p_i}{\varphi(A)}
\right)
\end{align}
The twist operator is Abelian \cite{govindarajan}
\begin{equation}
\mathcal F_\varphi = \exp\left\{
(N\otimes1)\ln\frac{\varphi(A\otimes 1+1\otimes A)}{\varphi(A\otimes 1)}
+(1\otimes N)\left(A\otimes1+ \ln\frac{\varphi(A\otimes 1+1\otimes A)}{\varphi(1\otimes A)} \right)
\right\}
\end{equation}
where $N=ix_i p^i$ and $[N,A]=0$.
Since this twist is Abelian, it automatically satisfies cocycle condition, and therefore it is a Drinfeld twist. Special case is presented in \cite{21, klry09}


\textbf{ii)} In the second case, leading to Jordanian twists, given by Borowiec and Pacho\l{} in \cite{Borowiec2008}, $\psi(A)$ is a linear function, i.e. $\psi(A)=1+rA$, where $r \in \mathbb R$, and $\gamma(A)=0$,
which leads to 
\begin{equation}
\varphi=\psi^{-\frac1r}=(1+rA)^{-\frac1r}
\end{equation}

The coproducts of momenta are:
\begin{align}
\Delta p_0 &= p_0 \otimes \varphi(A) + 1 \otimes p_0 \\
\Delta p_i &= p_i \otimes \psi(A) + 1 \otimes p_i
\end{align}
The family of corresponding twist operators is:
\begin{equation}
\mathcal F_r = \exp\left\{
\left(-{L^0}_0 + \frac1r{L^k}_k\right) \otimes \ln \varphi(A)
\right\}
\end{equation}
Special Jordanian twist was studied by Bu, Yee and Kim in \cite{22} and corresponds to $\mathcal S_1$, but with interchanged left and right side of the tensor product, and with $a_0 \rightarrow -a_0$ and $c=r+1$, i.e.
\begin{equation}
\mathcal F_r= \left.\tilde{\mathcal F}_{\mathcal S_1}\right\vert_{a_0\rightarrow-a_0, ~ c=r+1}
\end{equation}

\subsection{Light-cone deformation}\label{LCdef}

In the light-cone basis, the $\kappa$-Poincar\'e algebra was studied in \cite{KulishMudrov} and \cite{BorPach}, the corresponding twist is extended Jordanian twist, written in terms of two exponential factors. It is identical to transposed twist of $\mathcal F_{\mathcal C_4}$ with $a_\mu \rightarrow -a_\mu$, i.e. $\left.\tilde {\mathcal F}_{\mathcal C_4}\right\vert_{a_\mu \rightarrow -a_\mu}$. 

Extended Jordanian twist corresponding to light-cone deformation is:
\begin{equation}
\mathcal F_{LC}=e^{-iM_{+-}\otimes\ln\Pi_+}e^{-\frac i\kappa M_{+a}\otimes P^a\Pi^{-1}_+}
\end{equation}
where $[P_\mu, \hat x_\mu]=-i\eta_{\mu\nu}\left[1+(a\cdot P)\right]+ia_\mu P_\nu$, see equation \eqref{realizations}, and
\begin{equation}\begin{split}
&\Pi_+=1+\frac1\kappa P_+=1-a\cdot P \\
&a_0 = a_1 = \frac1{\sqrt2\kappa}, \quad a_j=0 \text{ for } j>1 \\
&P_\pm = \frac{P_0\pm P_1}{\sqrt2} \\
&M_{+-}=iM_{01}, \qquad M_{\pm j} = \frac i{\sqrt2}(M_{0j}\pm M_{1j}), \quad j>1
\end{split}\end{equation}
If we define
\begin{equation}\begin{split}
\mathcal A&\equiv -iM_{+-}\otimes\ln\Pi_+ \\
\mathcal B&\equiv -\frac i\kappa M_{+a}\otimes P^a\Pi^{-1}_+
\end{split}\end{equation}
then $[\mathcal A,\mathcal B]=\alpha \mathcal B$, where
\begin{equation}
\alpha\equiv1\otimes\ln\Pi_+
\end{equation}
and $\mathcal A$, $\mathcal B$ and $\alpha$ generate algebra \eqref{ABaalgebra}, given in Appendix \ref{AppC}. Using result \eqref{ResultC} from Appendix \ref{AppC}, it follows that $\mathcal F_{LC}$, written as one exponential function, is given by
\begin{equation}
\mathcal F_{LC}=\exp\left\{
-iM_{+-}\otimes\frac{(\Pi_+-1)\ln\Pi_+}{\Pi_+-1}-\frac i\kappa M_{+a}\otimes P^a \frac{\ln\Pi_+}{\Pi_+-1}
\right\}
\end{equation}
Using our notation, this result is
\begin{equation}
\mathcal F_{LC}= \exp\left\{
M_{\alpha\beta}\otimes a^\alpha P^\beta \frac{\ln(1-a\cdot P)}{a\cdot P}
\right\}
\end{equation}
which thus proves the relation
\begin{equation}
\mathcal F_{LC}= \left.\tilde {\mathcal F}_{\mathcal C_4}\right\vert_{a_\mu \rightarrow -a_\mu}.
\end{equation}
The twist $\mathcal F_{LC}$ leads to nonlinear realization \eqref{formula7A} and the corresponding coproduct is transposed coproduct with $a_\mu \rightarrow -a_\mu$.

\section{Outlook and discussion}

The full analysis of all possible linear realizations for $\kappa$-Minkowski space for time-, space- and light-like deformations is given. These realizations can be expressed in terms of the generators of $\mathfrak{gl}(n)$ algebra. Coproducts of momenta for linear realizations are constructed. We have presented a method for constructing Drinfeld twist operators corresponding to each linear realization of $\kappa$-Minkowski space and proved that it satisfies the cocycle and normalization conditions. We have constructed a whole new class of Drinfeld twists compatible with $\kappa$-Minkowski space and linear realizations, denoted by $\mathcal{C}_1$, $\mathcal{C}_2$, $\mathcal{C}_3$ and $\mathcal{C}_4$. The symmetries generated by Drinfeld twists are described by $\kappa$-deformed $\mathfrak{igl}(n)$-Hopf algebras, and in the special case of $\mathcal{S}_1$ and $\mathcal{C}_4$ we get the Poincar\'e-Weyl-Hopf algebra and light-like $\kappa$-Poincar\'e-Hopf algebra, respectively.   We further illustrate how our method  also works for constructing Drinfeld twists for nonlinear realizations and we compared our results to the examples already known in the literature. 

In this paper we were dealing mostly with linear realizations and the corresponding Drinfeld twists. However, for any realization, in general, one can construct a twist operator that does not have to satisfy the cocycle condition in the Hopf algebra sense (i.e. not a Drinfeld twist), rather  it satisfies the cocycle condition in a more general sense (up to  tensor exchange identities \cite{algebroid}), i.e. in the framework of Hopf algebroids \cite{algebroid}. It is crucial to notice that the $\kappa$-Minkowski space can be embedded into a Heisenberg algebra which has a natural Hopf algebroid structure. One can show that the  star product resulting from this generalized twist operator is  associative and that the corresponding symmetry algebra is a certain deformation of $\mathfrak{igl}(n)$-Hopf algebra. This general framework is more suitable to address the questions of quantum gravity \cite{3,4} and related  new effects to Planck scale physics.

The problem of finding all possible linear realizations is closely related to classification of bicovariant differential calculi on $\kappa$-Minkowski space \cite{forme}. Namely, the requirement that the differential calculus is bicovariant leads to finding all possible algebras between NC coordinates and NC one-forms that are closed (linear) in these NC one-forms. The corresponding equations for the structure constants (from the super-Jacobi identities) are exactly the same as eqs. (8,9). The linear realizations elaborated in this paper are expressed in terms of Heisenberg algebra, but one can extend this to super-Heisenberg algebra, by introducing Grassmann coordinates and momenta. This way one can construct the extended twists \cite{forme, Juric2012} which have the same desired properties, but also give the whole differential calculi.

With linear realization it is much easier to understand and to perform practical calculation in the NC space. In  \cite{beggs} it is proposed that the NC metric should be a central element of the whole differential algebra (generated by NC coordinates and NC one-forms). This NC metric should encode some of the main properties of the quantum theory of gravity. We hope that using the tool of linear realizations one can perform such calculations for a large class of deformations, and for all types of bicovariant differential calculi and predict new contributions to the physics of quantum black holes and the quantum origin of the cosmological constant \cite{MajidTao}.

Recently \cite{kratki}, the Drinfeld twist corresponding to $\mathcal{C}_4$ was analyzed and the corresponding scalar field theory was discussed. We are planing to further analyse the properties of quantum field theories \cite{ms11}, especially gauge theories that arise from this twist, but we are also interested in pursuing further investigations on the physical aspects of  $\mathcal{C}_{1,2,3}$ cases.

\section*{Acknowledgements}

Authors would like to thank S. Mignemi, A. Pacho\l{} and Z. \v Skoda for 
 comments.
This work has been fully supported by Croatian Science Foundation under the project (IP-2014-09-9582). 
The authors would also like to thank the anonymous referees for valuable remarks.

\appendix

\section{Derivation of coproduct $\Delta p_\mu$}
\label{AppA}
Here we present construction of equations for $K_\mu(k)$ and $P_\mu(k_1,k_2)$ and their solutions for linear realizations.

From equation \eqref{expPkk} we find
\begin{equation}
e^{-i\lambda k_1\cdot \hat x}p_\mu e^{i\lambda k_1\cdot \hat x} \triangleright e^{ik_2 \cdot x} =P_\mu(\lambda k_1,k_2)e^{ik_2\cdot x}
\end{equation}
where $(k_1)_\mu, (k_2)_\mu \in \mathbb M^n$ and $p_\mu \in \mathcal T$. Differentiating both sides by $\frac\partial{\partial\lambda}$ and using $\hat x_\mu = x^\alpha \varphi_{\alpha\mu}(p)$ we get the relation between $P_\mu(\lambda k_1,k_2)$ and realization $\varphi_{\mu\nu}(P(\lambda k_1,k_2))$:
\begin{equation}
\frac{\partial P_\mu(\lambda k_1,k_2)}{\partial\lambda}=\varphi_{\mu\alpha}(P(\lambda k_1,k_2))k_1^\alpha
\end{equation}
Note that for $\lambda=0$ 
the boundary condition is
\begin{equation}
P_\mu(0,k)=k_\mu. \label{Pbcon} 
\end{equation}

The coproduct for momentum $p_\mu$ is calculated by \cite{Meljanac09, Meljanac09a, Meljanac2011}:
\begin{equation}
\Delta p_\mu = \mathpzc D_\mu(p \otimes 1, 1 \otimes p)
\label{CoproductD} 
\end{equation}
where the function $\mathpzc D_\mu(k_1,k_2)$ is 
given by 
\begin{equation}
\mathpzc D_\mu (k_1,k_2) = P_\mu(K^{-1}(k_1),k_2)
\label{Ddef}
\end{equation}
and $K^{-1}_\mu(k_1)$  is inverse function of $P_\mu(k_1,0)=K_\mu(k_1)$.

The function $\varphi_{\mu\alpha}(p)$ describes the choice of realization in a following way:
\begin{equation}
\hat x_\mu = x^\alpha \varphi_{\alpha\mu}(p)
\end{equation}
In the case of linear realizations $\varphi_{\alpha\mu}(p)$ is:
\begin{equation}
\varphi_{\alpha\mu} = \eta_{\alpha\mu} + K_{\beta\mu\alpha}p^\beta 
\end{equation}
therefore
\begin{equation}
\frac{\partial P_\mu(\lambda k_1,k_2)}{\partial\lambda}=(k_1)_\mu + {K^\alpha}_{\beta\mu}P_\alpha(\lambda k_1,k_2) k_1^\beta
\end{equation}
This can be solved by expanding $P(\lambda k_1, k_2)$ in terms of $\lambda$.
\begin{equation}
P_\mu(\lambda k_1,k_2)=\sum_{n=0}^\infty P^{(n)}_\mu(k_1,k_2) \lambda^n,
\end{equation}
which, comparing the terms with the same power of $\lambda$, leads to
\begin{align}
&P^{(1)}_\mu(k_1,k_2)=(k_1)_\mu+{K^\alpha}_{\beta\mu}P^{(0)}_\alpha(k_1,k_2) k_1^\beta, \label{Pzero1} \\
&P^{(n+1)}_\mu(k_1,k_2)=\frac1{n+1}{K^\alpha}_{\beta\mu} P^{(n)}_\alpha(k_1,k_2) k_1^\beta \label{Pn1}, ~ \text{for }n\ge1.
\end{align}
Boundary condition \eqref{Pbcon} leads to:
\begin{equation}
P^{(0)}_\alpha(k_1,k_2)=(k_2)_\mu.
\label{Pzerosolution}
\end{equation}
For sake of brevity, let us define:
\begin{equation}
\EuScript K_{\mu\nu}(k) \equiv -K_{\mu\alpha\nu}k^\alpha.
\label{Kpdef}
\end{equation}
Using \eqref{Pzerosolution} and \eqref{Kpdef}, equations \eqref{Pzero1} and \eqref{Pn1} become:
\begin{align}
&P^{(1)}_\mu(k_1,k_2)=(k_1)_\mu- \EuScript K_{\alpha\mu}(k_1) k_2^\alpha, \\
&P^{(n+1)}_\mu(k_1,k_2)=-\frac1{n+1}{\EuScript K^\alpha}_\mu(k_1) P^{(n)}_\alpha(k_1,k_2), ~ \text{for }n\ge1.
\end{align}

The solution for $P_\mu(k_1,k_2)$ is:
\begin{equation}
P_\mu(k_1,k_2)=\left( \frac{\eta-e^{-\EuScript K(k_1)}}{\EuScript K(k_1)} \right)_{\alpha\mu} k_1^\alpha 
+ \left(e^{-\EuScript K(k_1)}\right)_{\alpha\mu} k_2^\alpha
\label{Psolution}
\end{equation}
Solution for $K_\mu(k)=P_\mu(k,0)$ is simply:
\begin{equation}
K_\mu(k)=\left( \frac{\eta-e^{-\EuScript K(k)}}{\EuScript K(k)} \right)_{\alpha\mu} k^\alpha 
\end{equation}
It is useful to define
\begin{equation}
k^W_\mu = K^{-1}_	\mu(k)
\label{pWdef}
\end{equation}
Inserting this definition and solution \eqref{Psolution} into \eqref{Ddef} we get
\begin{equation}
\mathpzc D_\mu (k_1,k_2) = P_\mu(k_1^W,k_2)=(k_1)_\mu+\left(e^{-\EuScript K(k_1^W)}\right)_{\alpha\mu}k_2^\alpha.
\label{SolutionD}
\end{equation}

The momentum $p^W_\mu=K^{-1}_\mu(p)$, introduced in section \ref{cofmomenta} (see eq. \eqref{pWact}), is related to $p_\mu$ via 
\begin{equation}
p_\mu=\left( \frac{\eta-e^{-\EuScript K}}{\EuScript K} \right)_{\alpha\mu} (p^W)^\alpha 
\label{ppW}
\end{equation}
and is given in closed form in \eqref{pwc1}, \eqref{pwc2}, \eqref{pwc3} and \eqref{pwc4} for the solutions $\mathcal C_1$, $\mathcal C_2$, $\mathcal C_3$ and $\mathcal C_4$ respectively. The momentum $p^W_\mu$ corresponds to Weyl symmetric ordering \cite{Meljanac-3,ms06,Meljanac-4}
\begin{equation}
[p^W_\mu,\hat x_\nu]=\eta_{\mu\nu} \frac{a\cdot p^W}{\text e^{-a\cdot p^W}-1}
+\frac{a_\nu p^W_\mu}{a\cdot p^W} \left(1+\frac{a\cdot p^W}{\text e^{-a\cdot p^W}-1}\right).
\label{pwhatx}
\end{equation}
For $p^W_\mu$, it is useful to define:
\begin{equation}
\EuScript K_{\mu\nu} \equiv \EuScript K_{\mu\nu}(p^W) = -K_{\mu\alpha\nu}(p^W)^\alpha.
\label{KpWdef}
\end{equation}
Using this definition, and solution \eqref{SolutionD} with the equation \eqref{CoproductD}, finally leads to the coproduct:
\begin{equation}
\Delta p_\mu = p_\mu\otimes1 + \left(e^{-\EuScript K}\right)_{\alpha\mu}\otimes p^\alpha
\end{equation}



\section{Construction of the twist operator from the coproduct of momenta}
\label{AppB}

It can be shown that for any $A_{\mu\nu}$ such that $[A_{\mu\nu},L_{\sigma\rho}]=0$ and $[A_{\mu\nu},A_{\sigma\rho}]=0$, the following holds:
\begin{equation}\label{noid}
:e^{{L_\beta}^\alpha {A_\alpha}^\beta}:=e^{{L_\beta}^\alpha{\left[\ln(1+A)\right]_\alpha}^\beta}
\end{equation}
This identity is a generalization of the result presented in \cite{ms06} 
 in equations (A. 16) and (A. 17). See also Section 2 in \cite{Meljanac2010}.

Twists can be calculated from the known coproducts of momenta using the equation \eqref{normaltwist}.
We would like to write the twist in the following form:
\begin{equation}
\mathcal F = e^f,
\end{equation}
where
\begin{equation}
f=\sum_{s=1}^\infty f_s,
\label{fexpansion}
\end{equation}
and $f_s \in \mathcal U\left[\mathfrak{igl}(n)\right] \otimes \mathcal U\left[\mathfrak{igl}(n)\right]$ is contribution to $f$ in $s$-th order of $\frac1\kappa$.

Inserting \eqref{alldeltap} into \eqref{normaltwist}, for $t=0$ we get:
\begin{equation}
\mathcal F^{-1}=:\exp\left\{{(\Lambda^{-1}-1)^\beta}_\alpha \otimes {L^\alpha}_\beta \right\}:
\end{equation}
From equation \eqref{noid} it follows
\begin{equation}
\mathcal F^{-1}=\exp\left\{ {(\ln\Lambda^{-1})^\beta}_\alpha \otimes {L^\alpha}_\beta \right\}
\end{equation}
Since $\Lambda_{\mu\nu}=\left(e^{\mathcal K}\right)_{\mu\nu}$, we find the twist: 
\begin{equation}
\mathcal F = \exp\left( \mathcal K_{\beta\alpha} \otimes L^{\alpha\beta} \right)
\end{equation}
Since $\mathcal K_{\mu\nu}=-K_{\mu\alpha\nu}(p^W)^\alpha$ and $\kl_\mu=-iK_{\beta\mu\alpha}L^{\alpha\beta}$, the result can also be written as:
\begin{equation}
\mathcal F=\exp\left(-ip^W_\alpha\otimes \kl^\alpha\right)
\end{equation}

\section{A special case of the BCH formula}
\label{AppC}
Let us consider algebra generated by $\mathcal A$, $\mathcal B$ and $\alpha$:
\begin{equation}\label{ABaalgebra}
[\mathcal A, \mathcal B]=\alpha \mathcal B, \qquad [\mathcal A,\alpha]=[\mathcal B,\alpha]=0
\end{equation}
then
\begin{equation} \label{ResultC}
e^{\mathcal A} e^{\mathcal B} = e^{\mathcal A+\mathcal Bf(\alpha)}
\end{equation}
where
\begin{equation} \label{falpha}
f(\alpha)=\frac\alpha{1-e^{-\alpha}}
\end{equation}

This can be proved by representing $\mathcal A$ and $\mathcal B$ with 2 by 2 matrices:
\begin{equation}
\mathcal A= \frac\alpha2 \begin{pmatrix} 1 & 0 \\ 0 & -1  \end{pmatrix}, \qquad
\mathcal B=\begin{pmatrix} 0 & 1 \\ 0 & 0  \end{pmatrix}
\end{equation}
These matrices satisfy the algebra \eqref{ABaalgebra}. Their exponentials are:
\begin{equation}
e^{\mathcal A}=  \begin{pmatrix} e^{\frac\alpha2} & 0 \\ 0 & e^{-\frac\alpha2}  \end{pmatrix}, \qquad
e^{\mathcal B}=1+\mathcal B=\begin{pmatrix} 1 & 1 \\ 0 & 1  \end{pmatrix}
\end{equation}
leading to
\begin{equation} \label{eAeB}
e^{\mathcal A} e^{\mathcal B} = \begin{pmatrix} e^{\frac\alpha2} & e^{\frac\alpha2} \\ 0 & e^{-{\frac\alpha2}}  \end{pmatrix}.
\end{equation}
On the other hand, since
\begin{equation}
\mathcal A+\mathcal Bf(\alpha) = \begin{pmatrix} \frac\alpha2 & f(\alpha) \\ 0 & -\frac\alpha2  \end{pmatrix}
\end{equation}
it follows that
\begin{equation} \label{eABfa}
e^{\mathcal A+\mathcal Bf(\alpha)}= \begin{pmatrix} e^{\frac\alpha2} & \frac{f(\alpha)}\alpha (e^{\frac\alpha2}-e^{-{\frac\alpha2}}) \\ 0 & e^{-{\frac\alpha2}}  \end{pmatrix}
\end{equation}
Comparing \eqref{eAeB} and \eqref{eABfa} gives $f(\alpha)$ in \eqref{falpha}.

\end{document}